# Nonlocal Metasurface Lens for Long-Wavelength Infrared Radiation

Federico De Luca[1], Sriram Guddala[1], Michele Cotrufo[1,2], Jimmy Touma[3], Adam Overvig[1,4] and Andrea Alù[1,5*]

[1]Photonics Initiative, Advanced Science Research Center at the Graduate Center of the City University of New York, New York, NY 10031, USA
[2]The Institute of Optics, University of Rochester, Rochester, NY 14627, USA
[3]Air Force Research Laboratory, Eglin, FL 32542, USA
[4]Department of Physics, Stevens Institute of Technology, Hoboken, NJ
[5]Physics Program, Graduate Center, City University of New York, New York, NY 10016, USA
*aalu@gc.cuny.edu

*Dielectric metasurfaces are structured thin films with thickness smaller than the wavelength that aim at replacing and enhancing conventional bulk optical components by structuring local resonances across an aperture. At visible and near-infrared frequencies, titania or silicon are routinely used as substrates to realize these ultrathin devices, ideally suited for conventional nanofabrication techniques. Unfortunately, directly scaling these design and material approaches to long-wave infrared frequencies is not practical, due to challenges in the required thicknesses and the presence of phonon absorption lines. Nonlocal metasurfaces based on extended resonances with a local geometric phase offer a compelling design platform that can address these challenges. They enable ultrathin metasurfaces, as they leverage lattice resonances, while they also offer multi-functionalities and frequency-selectivity, and they can be implemented in a range of low-loss material platforms. Here, we demonstrate nonlocal metalenses based on germanium thin films on a zinc-selenide substrate, operating around $10.3 \mu m$ within a deeply subwavelength device thickness of $1.45 \mu m$ (14% the free-space wavelength). We showcase a novel meta-unit geometry based on a square lattice with highly isotropic dispersion features, supporting a resonant geometric phase that is highly stable in frequency, simplifying the rational design of complex metasurface operations. The introduced platform promises highly multi-functional, low-profile meta-optics with enhanced meta-unit designs, compatible with the challenging thermal spectral region for imaging and sensing applications.*

**Introduction**

The intensity peak of blackbody radiation of objects at room temperature falls within the long-wave infrared region (LWIR, 8-12 µm), meaning that optical devices working in this spectral range are essential components for thermal imaging and sensing applications. However, the realization of lightweight and inexpensive optics for this wavelength range represents a cumbersome task, mainly because of the limited transferability to LWIR of traditional material platforms used for visible and short-wave infrared (SWIR) wavelengths. Indeed, mainstream oxide glasses and polymers are characterized by almost universal absorption across the LWIR, making them incompatible for this class of devices. Alternative materials, such as germanium, are typically used to realize bulky, expensive and heavy lenses operating in this spectral range.

The challenges posed by conventional LWIR devices' architectures can be addressed with metasurface flat optics, which promise enhanced and novel functionalities in considerably reduced volumes. A metasurface is a structured thin film with one dimension (along the out-of-plane direction) comparable to or smaller than the operating wavelength, and the in-plane geometry patterned with building blocks called 'meta-units', designed so that the collective optical response matches the desired functionality. Metalenses are capable of high-performance multi-functionality within a compact volume and are manufacturable with standard lithographic techniques. The geometrical features of the meta-units scale approximately with the desired operational wavelength, which introduces both advantages and challenges when designing and fabricating LWIR metasurfaces. Indeed, due to their larger in-plane feature sizes as compared to visible and near-infrared metasurfaces, LWIR metalenses are easier to fabricate with standard nanofabrication techniques. In particular, their feature sizes are compatible with optical lithographic techniques, which, contrary to electron-based lithography, can be scaled to large-area manufacturing. On the other hand, as also the out-of-plane dimension increases with the operational wavelength, applying standard metalens designs to the LWIR range results in thick films, of the order of 10 microns, increasing the challenges in achieving high-quality devices. A metalens design platform with a thickness comparable to the one used at near-infrared frequencies (i.e., around 1 micron), is highly desirable in practice.

In conventional metasurface designs, the unit cells typically support a local resonant optical response, where 'local' means that nearest neighbor effects are ignored in the design stage, and each meta-unit is assumed to behave independently in shaping the optical wavefront. Nevertheless, while promising in terms of size weight and power (SWaP) improvements, local metalenses offer minimal control over the spectral response. Recently, nonlocal metasurfaces based on engineered quasi-bound states in the continuum (q-BICs) have been introduced to realize frequency-selective wavefront shaping [1]. Due to their design rooted in symmetry-breaking, the approach is exportable to a wide range of frequencies and material platforms, and can realize ultrathin lenses on the order of 15% of the free-space wavelength [4]. The spatial dispersion of these devices offers new opportunities for selective responses in metasurfaces [5], including simultaneous polarization, spatial and spectral selectivity, promising for applications such as cascaded meta-optics [4], augmented reality [4, 6, 7] and thermal emission control [8]. While frequency-selective (but spatially non-selective) metasurfaces based on localized high quality-factor ("high-Q") resonances [9] have been realized with both rectangular [10] and square lattices [11, 12], so far the dominant method for nonlocal metasurface lenses has relied on a rectangular lattice of two elliptical dimers that are rotated to impart a geometric phase, originally reported in [4] and widely adopted afterwards [13-22]. However, the anisotropic lattice results in spatial dispersion that differs in the two in-plane directions, giving a distinct nonlocality (and therefore, selectivity) depending on the direction. For wavelength and polarization selective radial lensing, quasi-isotropic dispersion associated with square and hexagonal lattices is an important yet missing tool for nonlocal metalenses. Another shortcoming of the rectangular lattice has been the dependence of the q-BIC resonant frequency on the orientation angle used to impart a phase profile. This complication is primarily born of the low symmetry of the rectangular lattice controlling the q-BIC; recently, q-BICs with resonant wavelengths that are stable with the strength of geometric perturbation have been reported [23, 24], but a design with stability regardless of geometric phase has so far remained elusive.

In this work, we design and experimentally realize nonlocal metalenses based on germanium thin films on a zinc selenide substrate, which, by operating at 10.3 µm, enables spectral and polarization selective lensing in the LWIR with a device thickness of 1.45 um, corresponding to 14% the free-space wavelength. We also introduce a novel meta-unit geometry relying on a square lattice characterized by a highly isotropic dispersion and a resonant geometric phase that requires no frequency adjustment. This approach simplifies the design compared to conventional dimer approaches [13-22], and it can be adopted for visible and near-infrared nonlocal metasurfaces. The proposed platform paves the way to highly multi-functional, low-profile meta-optics for the thermal spectral region.

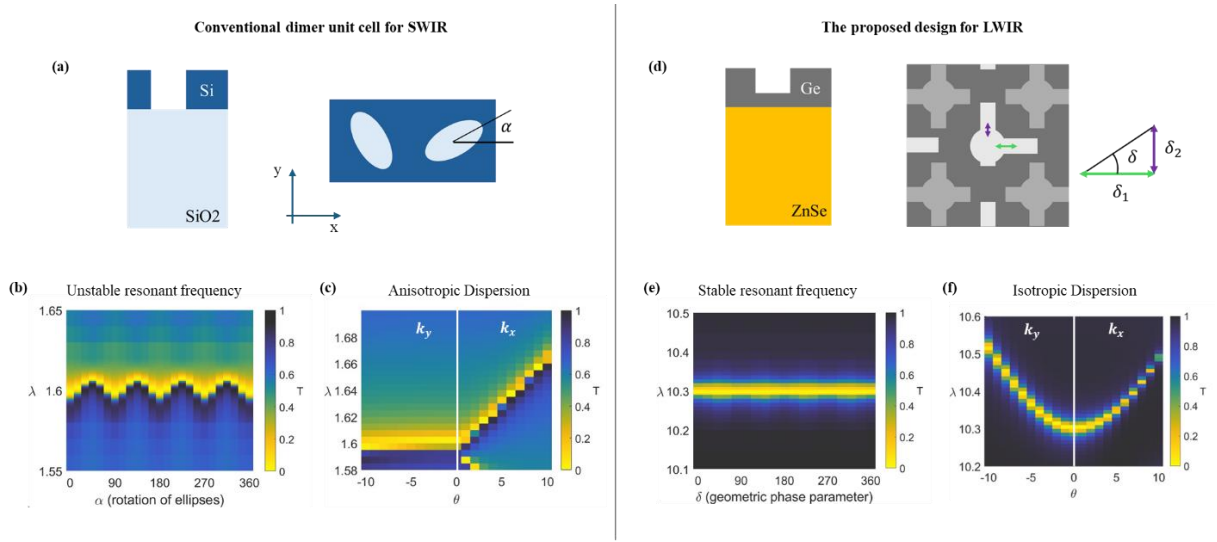

**Figure 1. Novel meta-unit geometry characterized by a square lattice.** The proposed design for LWIR (right panel), based on a thin film of germanium on top of a zinc selenide substrate, is compared to a conventional silicon on silica dimer unit cell for the SWIR (left panel). (a) Schematics of the conventional dimer defining the in-plane orientation angle α, the parameter that controls the resonant phase of the meta-unit. (b-c) Transmittance as function of wavelength λ and α (plot b), and of λ and angle of incidence θ (plot c). (d) The proposed design is characterized by a constant lattice of "unperturbed" crosses, while at the interstitial sites there are perturbed crosses that spatially vary. The direction δ, given by the sum of horizontal and vertical displacements of the rods, governs the symmetries of the q-BIC and its vectorial properties. (e-f) Transmittance as function of wavelength λ and δ (plot e), and of λ and angle of incidence θ (plot f) showing that, conversely to the conventional geometry, the novel design has highly isotropic dispersion while supporting a resonant geometric phase that requires no frequency adjustments.

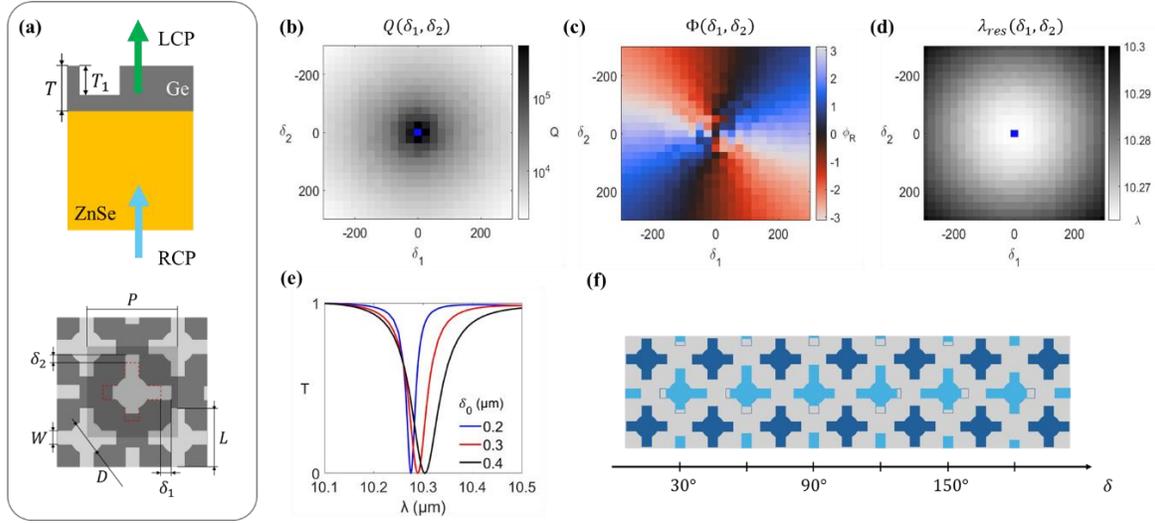

**Figure 2. The proposed metasurface operates in transmission, transforming right-handed to left-handed circular polarization and focusing the converted signal.** (a) Schematics and characteristic parameters of the proposed meta-unit. A resonance at about 10.3μm is obtained with the following parameters: P = 3.15μm, L = 2.1μm, W=0.65μm, D=1.1μm. (b-d) Q-factor, resonant phase Φ, resonant wavelength λ$_{res}$, respectively, as a function of δ$_1$ and δ$_2$. (e) Transmittance as a function of wavelength λ for three distinct values of $\delta_0 = \sqrt{\delta_1^2 + \delta_2^2}$, which allows to control the Q-factor. δ$_0$=0.4μm in this work. (f) Schematics showing how the positions of the rods of the perturbed crosses define δ.

## Results - Design

The working mechanism of our nonlocal metalens is rooted in the physics of perturbative nonlocal diffractive metasurfaces sustaining a q-BIC in high-index contrast photonic crystal slabs (PCS), which are an emerging technology for realizing compact, highly tailored optical devices [1]. Bound states in the continuum (BICs) are non-radiating confined resonances coupled to free space with an unbounded Q-factor. They can emerge due to symmetry protection, or due to reasons unrelated to it. Symmetry-protected BICs become radiative if the symmetry of the PCS is reduced, i.e., if the lattice of the PCS is perturbed such as to induce a net dipole moment that can couple to free space. The result is a q-BIC whose optical lifetime properties, expressed via the Q-factor, are controllable by the means of a perturbation δ, where $Q \propto \frac{1}{\delta^2}$ [2,3]. The vectorial polarization properties of q-BICs are determined by the type of perturbation, making use of specific selection rules catalogued in Ref. [2].

The symmetries of the metasurface lattice, the shape and orientation of the unit cells and the employed material platform are crucial parameters for defining the functionalities of nonlocal metasurfaces. In Fig. 1 we compare our design strategy with a conventional rectangular dimer unit cell for SWIR, made of a thin film of silicon on silica. The design we propose is characterized by a constant lattice of "unperturbed" crosses, while at the interstitial sites we find perturbed crosses that spatially vary. The perturbation is realized by displacing the horizontal rod along the *x* axis by $\delta_1$, and/or the vertical one along the *y* axis by $\delta_2$, with respect to the center of the circles. Here, the orientation $\delta = atan2(\delta_2, \delta_1)$ of the net dipole, given by the sum of horizontal and vertical dipoles, governs the symmetries of the q-BIC and its vectorial properties. To extend the range of applicability of nonlocal metasurfaces from the visible and SWIR to the LWIR range, we employ thin films of amorphous germanium (n ~ 4.40) on top of a zinc selenide (n = 2.40) substrate, that enable low losses and high index contrast between the film and the substrate. Titania and silica, standard materials at visible and SWIR frequencies, cannot be used in the LWIR because of their phonon absorption lines. High-quality crystalline silicon can have sufficiently low loss for local, non-resonant metasurface devices [25], but amorphous silicon is not suitable due to high absorption losses. The typical meta-unit of conventional nonlocal metasurfaces is a dimer, whose orientation, described by the angle α in Fig. 1a, controls the resonant geometric phase. Nevertheless, as shown in Fig.1b, where the normal-incidence transmittance of a periodic metasurface made of dimers is reported as a function of α and of the wavelength, the resonant wavelength of the q-BIC depends on α, thus requiring a more complicated meta-unit library to compensate for the shift in resonance and keep all the meta-units resonant at the same wavelength and Q-factor [3,4]. Moreover, as shown in Fig. 1c, while the resonant wavelength of the nonlocal mode is dispersive with the angle of incidence θ, the two bands, for oblique incidence along *x* or *y* respectively, are remarkably asymmetric. The curvature and dispersion are intimately tied to the selectivity of the response of nonlocal devices, yielding distinct responses in each of the lattice directions, reminiscent of astigmatism [14]. Our simplified square lattice and geometric perturbation eliminates both challenges to yield highly wavelength-stable q-BICs (Fig. 1e) with near isotropic dispersion (Fig. 1f).

Our design consists of an array of partially etched holes in a film of germanium, as described in Fig. 2a. The thicknesses T and $T_1$ are optimized to have large spectral separation between the multiple q-BIC modes that arise from the broken symmetry [2-4] to maximize the transparency of the device outside the resonant mode, and to introduce out-of-plane symmetry breaking for the sake of improved conversion efficiency. The purpose is to allow a clear experimental demonstration of the contrast between device performance on and off resonance, while minimizing the overall metalens thickness and maintaining broadband transmittance [2-4]. The characteristic meta-unit parameters are also reported in Fig. 2a. In this design the Q-factor $Q$, resonant phase $\Phi$ and resonant wavelength $\lambda_r$, can be controlled by varying $\delta_1$ and $\delta_2$ (Fig. 2b-d). In particular, the geometric phase can be spatially varied while maintaining Q and $\lambda_r$ at a constant value by adjusting the perturbation along a circular contour in Fig. 2b-c. In other words, the optical lifetime of the q-BIC and resonant wavelength can be defined by fixing $\delta_0 = \sqrt{\delta_1^2 + \delta_2^2}$, as shown in Fig.2e, where transmittance is reported for three distinct values of $\delta_0$, and $Q \propto \frac{1}{\delta_0^2}$. Then, the angle δ governs the symmetries of the q-BIC and its vectorial properties (see Fig. 2f and Appendix). As δ varies continuously from 0 to 2π, the resonant polarization angle must vary accordingly, i.e., the resulting resonant phase is $\phi_r \sim \delta$. In other words, the meta-unit behaves like a dichroic optical element characterized by the orientation of the dimerizing perturbation δ.

For circularly polarized incident light matching the q-BIC resonance, two projections of the polarization basis are necessary to analyze the outgoing signal [2-4]. First, when coupling into the element, the input signal is decomposed into two linear polarizations, parallel or perpendicular to $\phi_r$, which are resonantly reflected or transmitted, respectively. Second, the outcoupling corresponds to a decomposition of the signals into their constituent spins, i.e. right- (RCP) or left- (LCP) circularly polarized light. Each of the two steps halves the energy of the signal and the result is a four-port system with ~25% of the incident power sent to each of the ports, corresponding to transmitted or reflected light for each of the circular polarizations [1]. For a given input circular polarization, a geometric phase is imparted to the transmitted signal of converted (i.e., opposite) handedness and to the reflected signal of preserved handedness only at the narrowband q-BIC resonances. For instance, in the case of incoming RCP light, as in this work, a geometric phase Φ= 2ϕ ~ 2δ is imparted into the transmitted LCP signal and the reflected RCP signal. A 2D spatial distribution of the dimerizing perturbation δ thus encodes a 2D geometric phase profile. This geometric phase factor of 2 represents a further distinction from the dimer approach in Fig. 1, which follows $\Phi = 2\phi \sim 4\alpha$ due to the *p2* symmetry of the lattice, i.e. the meta-units defined by $\alpha$ and $\alpha + 180°$ are identical geometries, meaning $\phi \sim 2\alpha$. Here, the *p1* symmetry (see Appendix) implies that we require an evolution of δ by 360° to return to an identical meta-unit, meaning $\phi \sim \delta$.

**Results - Characterization**

We designed and fabricated (Fig.3) a square-shaped nonlocal metalens with side length equal to 900μm and a hyperboloidal geometric phase profile [26]. The reflectance spectrum of the metasurface (Fig. 3a) was measured using a commercial Fourier transform infrared (FTIR) microscope. Due to technical limitations of the microscope, these measurements were performed with unpolarized impinging light. The spectrum in Fig. 3a exhibits a Fano like resonance at the designed wavelength near 10.3μm, with a Q-factor on the order of 100. Even if the peak is significantly smoothed, likely due to a reduced signal-to-noise ratio associated with unpolarized light incidence, the spectrum provides valuable information and a reference for the analysis of the focusing properties. The metalens is designed to perform conversion from RCP to LCP and to focus the converted signal. Thus, to perform a full analysis of the device response, two quarter waveplates (QWP), one in the input path (to generate a CP input) and one in the output path (to select a certain CP state), would be required. However, the availability of QWPs in the LWIR is currently limited, and the few readily obtainable are very expensive. This not only highlights the importance of developing new technologies for the LWIR band, but it also motivates us to use, for a complete characterization of the properties and the performance of our metalens, two distinct and complementary configurations of the same custom-built transmission-mode microscope employing a single QWP (see Methods). In one configuration, shown in Fig. 3c, the input signal is CP while the outcoming signal is not. Dually, in the CP resolved configuration represented in Fig. 3g, the input is linearly polarized, while the transmitted signal is collected by the camera after being filtered with the QWP. In both cases, the metalens is mounted on a linear translational stage moving along the *z* direction.

A series of transverse 2D far-field scans recorded for RCP incidence, reported in Fig. 3d, shows that a focal spot can be observed within a bandwidth of ~ 400 nm, from 10.20 µm to 10.60 µm, with peak intensity at the resonant wavelength of 10.31 µm. As expected, a similar operation is not observed for LCP incidence (which operates as a diverging lens, as usual for geometric phase metasurfaces). As an indirect measure of the polarization of the output, we also measured and compared the normalized integrated field intensities, $\eta_F = \frac{I_F}{I_T}$ and $\eta_B = \frac{I_B}{I_T}$, at resonance as a function of the rotation angle of a linear polarizer placed in front of the imaging camera (see plots in Fig. 3e-f). Here, $I_F$ is the integrated field intensity within the FWHM of the focal spot, while $I_B$ is field intensity integrated in the complementary region outside the focal area, i.e. in the background area. These quantities are normalized to the total intensity ($I_T = I_F + I_B$). The maximum of the integrated focal spot intensity is 17.5% of the total signal (focused and unfocused). The polar plots in Figs. 3e-f show that the polarization ellipse for focused light intensity (plot e), is phase-shifted 90° relative to that of the unfocused signal (plot f). Such phase shift indicates, as expected from theory, that the focused signal has perpendicular polarization relative to that of the unfocused signal, supposedly coming from unconverted incident RCP beam. However, a final proof requires to resolve the output with a QWP.

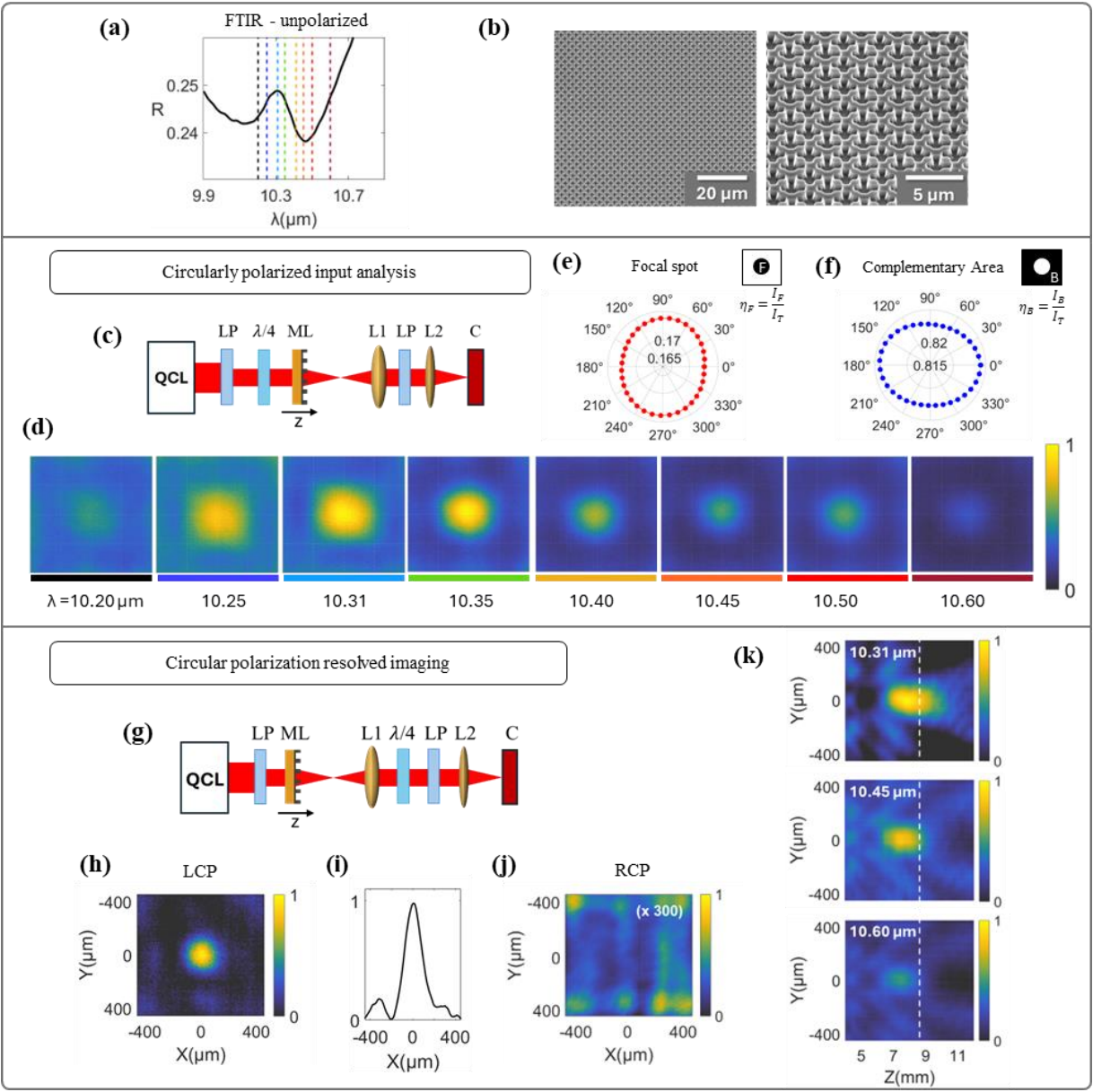

**Figure 3. Characterization of a square nonlocal metalens with a side length equal to 900μm and hyperboloidal geometric phase profile.** (a) Reflectance spectrum of the metalens measured using a FTIR microscope under unpolarized light incidence. Dashed lines correspond to scanning wavelengths of plot d. (b) SEM images of the fabricated sample. The central part (panels c-f) reports a schematic of the setup (plot c) and data (plot d-f) when the input signal is circularly polarized while the imaged signal is linearly polarized. (d) Transverse 2D far-field scans recorded for RCP incidence. (e-f) Normalized integrated field intensities, $\eta_F$ ans $\eta_B$ respectively, at resonance as a function of the rotation angle of a linear polarizer placed in front of the imaging camera. The polarization ellipse in plot (e) is derived from integrating within the FWHM of the focal spot, while plot f is obtained integrating in the complementary region outside the focal area, i.e. the background area. In both cases the integrated intensities, $I_F$ and $I_B$, are normalized to the total intensity ($I_T = I_F + I_B$); the maximum focal spot intensity measured is 17.5% of the total intensity. The bottom part (panels g-k) analyzes the scenario in which linearly polarized light impinges on the metasurface while the imaged signal is circular polarization resolved (plot g). (h) 2D far-field scan for LCP resolved output measured at resonance and (i) corresponding 1D linecut in correspondence of the intensity peak. (j) Saturated 2D far-field scan for RCP resolved output measured at resonance. (i) Longitudinal 2D far-field scans at different wavelengths, obtained recording the focal spot as a function of the position of the metalens along the z-axis, revealing that the focal spots at resonance ($\lambda = 10.31$ μm) are much brighter than the focal spots off resonance.

Next, we investigated the case where the metasurface is excited by a linearly polarized excitation, and the CP state of the transmitted light is analyzed. Since a linearly polarized input is the sum of two perpendicular circular polarization, the signal transmitted by the metalens has four components: RCP unconverted, RCP converted to LCP, LCP unconverted, and LCP converted to RCP. By adding in the output path a polarization analyzer, formed by a linear polarizer and a QWP, we can selectively remove two of these components. The metalens is expected to convert impinging RCP into a focused LCP spot, while it does not affect impinging LCP. Thus, by setting the polarization analyzer to collect LCP light, we expect to be able to see a spot created by the RCP component of the excitation, while the non-converted LCP light will provide a much weaker background. The measurement in Fig. 3h confirms this expectation, as it shows a clear focusing on resonance and when LCP is selected in the output path. Instead, when the collected polarization is RCP, no focal spot is observed, as shown in Fig. 3j. The beam waist (FWHM) is approximately 200 μm (Fig.3i). Chromatic aberration can be assessed by analyzing Fig.3k, which shows a series of longitudinal 2D far-field scans at different wavelengths, obtained recording the focal spot as a function of the position of the metalens along the z-axis. This scan confirms that the focal spots at resonance (λ = 10.31 μm) are much brighter than the focal spots off resonance. The chromatic aberration, corresponding to the focal shift as a function of wavelength, is found to be less than 500 μm (dashed line in Fig.3i). Finally, we can estimate the metalens conversion efficiency via the ratio $\rho = \frac{I_{LCP}}{I_I}$, where $I_{LCP}$ is the integrated intensity of the total LCP output signal and $I_I = I_R + I_L$ is the total intensity impinging on the metalens. Here, $I_R$ and $I_L$ are the RCP and LCP integrated input intensities, respectively, measured by removing the metalens from the setup in Fig.3a. From the data in Fig. 3, we obtain $\rho \approx 12\%$. Note however that this value overestimates the actual conversion efficiency because $I_{LCP}$ includes also the unconverted background LCP signal. A better estimation can be done by considering only the intensity within the focal spot. Then, since the calculated focal spot intensity is about 17.5% of the total intensity, the ratio $\rho$ reduces to ~2%. Finally, by remembering that, under linear polarization input, half of the intensity (the LCP component) is wasted, we can estimate the metalens RCP-to-LCP focusing efficiency to be ~4%.

**Conclusions**

In this work, we demonstrated a nonlocal metalense based on germanium thin films on a zinc selenide substrate, operating near 10.3μm with a device thickness of 1.45μm (14% the free space wavelength). We showcased a novel meta-unit geometry based on a square lattice with highly isotropic dispersion while supporting a resonant geometric phase that requires no frequency adjustments. Our approach is compatible with cascaded nonlocal optics for the first time at LWIR. The nonlocality/spectral selectivity has important implications for the LWIR due to "chemical finger printing" of many molecules — our devices can act as both filter and lens in one, ultra-compact device, isolating and imaging target gases, for example. However, even if demonstrated at LWIR due to the unique need for ultrathin devices limited to specific materials, our design is rooted in symmetry, so it can be extended to any wavelength and could also be multi-layer to induce intrinsic chirality and higher efficiency [13]. In addition, the isotropic response in our system opens pathways for customizing the dispersion in nonlocal metasurfaces, which has been limited largely to rectangular lattices and therefore anisotropic responses best suited for cylindrical lenses rather than the radial lenses shown here. What is more, the simplicity of not separately adjusting Q and $\lambda_r$ as we vary polarization $\phi_r$ is a key advantage enabled by the higher symmetry of our design compared to the dimer approach. This may open pathways for nonlocal metasurfaces at SWIR and VIS, as well as thermal emission engineering, photoluminescence, emission from quantum and 2D materials, and nonlinear optics. Finally, the distinct manifestation of the geometric phase is consistent with recent results showing generalized Pancharatnam-Berry phases [27], suggesting nonlocal metasurfaces should, in the future, explore distinct lattice symmetries beyond rectangular (past) and square (here) to broaden the flexibility of the toolbox for photonics engineers. This suggests new work in hexagonal lattices, which may further improve the isotropy of the response due to 6-fold symmetry, while also representing a novel example of a generalized geometric phase.

## APPENDIX

**Classification according to Selection Rules for q-BICs.**

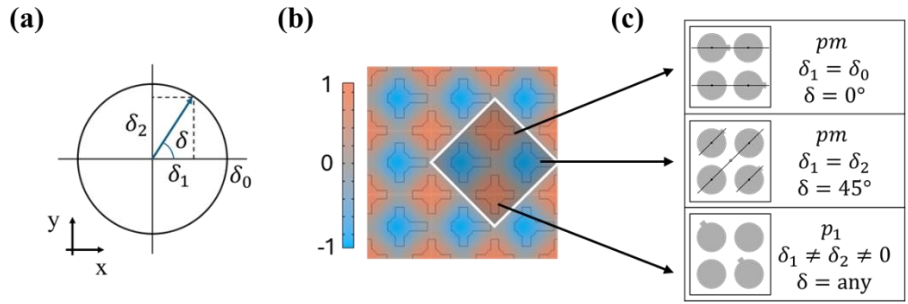

**Figure 4.** (a) Schematic of the relationship between perturbation parameters of the meta-unit introduced in Fig. 1d. Here, it can be clearly seen that $\delta_0 = \sqrt{\delta_1^2 + \delta_2^2}$, which means that the net perturbation, defining the Q-factor of the q-BIC, is constant. The graphic shows also that we require an evolution of $\delta$ by 360° to return to an identical meta-unit, meaning that the resonant phase $\phi \sim \delta$. (b) Out-of-plane component of the magnetic field in correspondence of the q-BIC resonance of Fig. 1e. The shadowed square area is taken into account for defining the irreducible representation of the mode and classifying its symmetry groups in plot c, according to the Selections Rules of Ref. [2]

Our meta-unit has in general the symmetries of the p1 space group, a lattice with no fold symmetries except for specific cases. The q-BIC under study is a TE mode at the high symmetry point X of a square lattice. It has the mode characteristics of the irreducible representation B2 in the first order Brillouin zone [2]. For specific combinations of $\delta_1$ and $\delta_2$, the meta-unit has the symmetries of two "parent" pm space groups with higher symmetry, which are combined to build the "child" p1 space group of lower symmetry. Indeed, pm groups have a reflection axis along a specific direction and no rotations or glide reflections and, if perturbations corresponding to pm groups with different reflection axis are added successively, the fold symmetry is broken resulting in a p1 space group, that allows only translations. The pm parent space group in the first row ($\delta_1 = \delta_0$ and $\delta_2 = 0$) allows coupling to a field polarization parallel to the x axis (that is, x polarization), while the other parent pm space group ($\delta_1 = \delta_2$) to a polarization angle of 45° from the x axis. The child space group p1 can then be parameterized by the angle $\delta$ that yields one pm space group when $\delta = 0°$ and the other pm group when $\delta = 45°$.

## METHODS

### Numerical Simulations

The plots in Fig. 1 and Fig. 2 are full-wave simulations of a periodic metasurface. The study employs the finite-element method via the commercial software COMSOL Multiphysics.

### Device fabrication

The metalens was fabricated employing a standard top-down lithographic process. Ø1/2" ZnSe Broadband Precision Windows (Thorlabs Inc.) were used as transparent substrates. These were cleaned by placing them in an acetone bath inside an ultrasonic cleaner, and later in an oxygen-based cleaning plasma (PVA Tepla IoN 40). After cleaning, a layer of 1.45 um of amorphous Germanium was deposited via electron-beam evaporation (AJA International Orion-8E). The sample was then spin-coated with a layer of electron-beam resist (ZEP 520-A) for 45 s at 3000 RPM and baked at 180°C for 5 min. The device was patterned with an electron beam lithography tool (Elionix ELS-HS50, 50 keV) at a current of 2 nA and a dose of 200 uC/cm$^2$. After exposure, the ZEP was developed by immersing the sample in chilled n-amyl acetate for 50 s and isopropyl alcohol for 15 s. Next, the exposed area was vertically etched to a depth of 0.87 um via a $C_4F_8/SF_6$ dry etching in an inductively coupled plasma machine (Oxford PlasmaPro System 100). The parameters of the optimized etching recipe are $C_4F_8$ 47sccm, $SF_6$ 14sccm, $O_2$ 15sccm, pressure 8mTorr RF Forward power 10W, ICP Forward power 1000W, leading to an etching rate of about 2.5 nm/s at 5°C table temperature. The resist mask was finally removed by immersing the sample in an N-methylpyrrolidone (NMP - Remover PG) solution heated to 180 °C for 10 minutes.

### Optical Characterization

The reflectance of the sample was measured with a Fourier transform infrared microscope (Bruker Invenio R) using a refractive ZnSe objective with numerical aperture NA = 0.2, and by normalizing the sample spectral response for unpolarized light incidence with respect to the reflection spectrum of a gold mirror.

The 2D far-field scans in Fig.3 were recorded with a custom-built transmission mode microscope employing a Hedgehog™ mid-infrared quantum cascade laser (QCL), a tunable quarter-wave plate (QWP) (Alphalas PO-TWP-L4-25-FIR), two ZnSe planoconvex lenses ($f$ = 5cm and $f$ = 15cm,) and a thermal camera (WinCamD-IR-BB, DataRay), arranging in two distinct configurations as shown in Fig. 3-4. The incident laser beam was collimated to the sample size of ~1mm. For the metalens circular polarization incidence performance, a linear polarizer and a tunable QWP are used to generate circular polarized light incidence into the metasurface, successively the transmitted focused/unfocused signal after the metalens is collected using an objective ZnSe lens with a 5cm focal length and another ZnSe plano convex lens with 15cm focal length as a tube lens in front of the thermal camera. For the CP resolved measurements, metasurface are imaged under linearly polarization incidence, while the focused and unfocused transmitted signal is collected by the camera after being resolved into two opposite handedness circular polarizations using a combination of tunable QWP and a linear polarizer inserted before the camera. For the metalens chromatic aberration and focal length measurements, the metalens focus was imaged on thermal imaging camera at different wavelengths, about the resonance wavelength of 10.31mm, by translating the metalens along the z- direction with a step resolution of 0.1mm.